\documentclass[prb,twocolumn,showpacs]{revtex4}
\usepackage{amssymb}
\usepackage{graphicx}
\usepackage{amsfonts}
\usepackage{amsmath}

\begin{document}

\title{Spin-torque generation by dc or ac voltages in magnetic layered structures}

\author{F. Romeo$^{1}$, and R. Citro$^{1,2}$}
\affiliation{$^{1}$Dipartimento di Fisica ''E. R. Caianiello'' and
C.N.I.S.M., Universit{\`a} degli Studi di Salerno, Via S. Allende,
I-84081 Baronissi (Sa), Italy\\
$^{2}$Laboratorio Regionale INFM-CNR SuperMat, Via S. Allende,
I-84081 Baronissi (Sa), Italy }

\begin{abstract}
A general expression of the current induced spin torque in a
magnetic layered structure in the presence of external dc or ac
voltages is derived in the framework of the scattering matrix
approach. A detailed analysis is performed for a
magnetic-nonmagnetic-magnetic trilayer connected to external leads
in the presence of dc voltage bias in the ballistic regime.
Alternatively, the possibility of producing spin torque by means
of the adiabatic ac modulation of external gate voltages (quantum
pumping) is proposed and discussed.
\end{abstract}

\pacs{73.23.-b,72.25.Pn,75.60.Jk,72.15.Qm}
%73.23.-b electron transport in mesoscopic system
%72.25.Pn spin pump current-driven
%75.60.Jk magnetization reversal
%72.15.Qm scattering in electron transport metals

\keywords{spin-torque, spin pumping, scattering matrix}

\maketitle

\section{Introduction}

Multilayers of alternating magnetic (generally ferromagnetic) and
nonmagnetic metal layers have recently attracted a lot of
attention because of giant magnetoresistance (GMR) effects. In
fact their electrical resistance depends strongly on whether the
moments of adjacent magnetic layers are parallel or antiparallel
and this effect has allowed the development of new kinds of
magnetic memory devices.\cite{memory_devices} The origin of the
GMR effect is the stronger scattering of conduction electrons by a
magnetic layer when their spins lie antiparallel to the layer's
magnetic moment compared to the case when their spins are parallel
to the moment. Thus the orientations of magnetic moments can
affect the flow of electrons, but as a reciprocal effect a
polarized electron current scattering from a magnetic layer can
affect the moment of the layer itself. In fact, as proposed by
Berger\cite{berger_spintorque} and
Slonczewski,\cite{slonc_spintorque} an electric current passing
perpendicularly through a magnetic multilayer may exert a torque
on the moments of the magnetic layers. This effect is known as
"spin transfer'', and may alter the magnetization state of the
layer. Of course this is a different mechanism from the effects of
current induced magnetic fields and offers the possibility of
realizing new kinds of magnetic devices. Importantly, it can serve
as a mean to detect the spin dynamics at nanometer scale and
measure spin currents. However, in order to utilize these effects
in real devices, it is necessary to achieve a quantitative
understanding of spin-current-induced torques and also provide
different ways for their generation.

In the present literature a semiclassical WKB approximation with
spin dependent potentials in 1D has been proposed in
Ref.[\onlinecite{slonc_spintorque}]. A generalization of this
study has appeared in Ref.[\onlinecite{peeters2008}] and an
extension of this calculation to take into account the band
structure effects on the degree to which an electron is
transmitted through a magnetic/nonmagnetic interface has been
proposed by Brataas et al.\cite{brataas_spintorque_kineq} by means
of the kinetic equations for spin currents and by Waintal et
al.\cite{waintal2000} by using a Landauer-B\"{u}ttiker type of
approach. A first principle theory based on the spin mixing
conductance appeared in [\onlinecite{carva_2007}].

Here we present a general formalism based on the scattering matrix
approach to calculate the torques in a magnetic layered structure
in the ballistic regime. Then we use this formalism to make an
explicit calculation for a magnetic(M)-nonmagnetic(NM)-magnetic(M)
trilayer connected to two metallic external leads. In this case
the ballistic regime is characterized by a spin-diffusion length
$l_s$ and a mean-free-path $l_m$ larger than the NM spacer. We
perform first the calculation in the presence of a dc voltage bias
applied to the leads, or equivalently for currents perpendicular
to the microstructure, then we proposed an alternative way of
achieving a spin-torque by the adiabatic ac modulation of two
system parameters (in our case the barrier heights at the M-NM
interface), i.e. by a quantum pump mechanism\cite{thouless}.
Although we limit our calculation to the single channel case,
differently from Ref.\onlinecite{waintal2000} all spin-scattering
processes at the M-NM interface are properly taken into account,
including spin-flip mechanisms.

The organization of the paper is the following: In
Sec.\ref{sec:model} we introduce the model Hamiltonian and the
general expression for the spin-torque. In
Sec.\ref{sec:scattering_formalism} we develop the scattering
matrix approach generalized for the calculation of the spin-torque
in a magnetic layered structure. In Sec.\ref{sec:pumping} we
discuss the adiabatic quantum pumping of spin-torque and finally
in Sec.\ref{sec:calculation} we present the results for a specific
system, a trylayer of M-NM-M metals connected to two nonmagnetic
leads. A quantitative analysis of the spin-torque components is
presented both in the case of a dc bias applied to the external
leads and when a quantum pump mechanism is activated. The two ways
of generating a spin-torque presents some differences that are
comparatively discussed. These differences could be important for
the realization of new magnetic devices.

\section{ The model and formalism}
\label{sec:model}

Let us consider a 1D system whose Hamiltonian is:
$H=-\frac{\hbar^2}{2m}\partial^2_x+V(x,\vec{\sigma})$, where
$V(x,\vec{\sigma})$ is some spin-dependent potential operator. The
field operator describing the electron state $\Psi$, obeys the
Schr\"{o}dinger equation $\overrightarrow{H} \Psi=i \hbar
\partial_t \Psi$ and $\Psi^{\dag}\overleftarrow{H} =-i \hbar
\partial_t \Psi^{\dag}$ and the time evolution of the
electron charge density $\rho=\Psi^{\dag}e\Psi$ is determined by
the Heisenberg equation:
\begin{eqnarray}
\label{eq:heisenberg}
\partial_t(\Psi^{\dag}e\Psi)=\frac{ie}{\hbar}[\Psi^{\dag}(\overleftarrow{H}-\overrightarrow{H})\Psi]=-\partial_x J,
\end{eqnarray}
where $J=\Psi^{\dag}e\hat{v}\Psi$ is the current density, while
$\hat{v}=\frac{i\hbar
}{2m}(\overleftarrow{\partial}_x-\overrightarrow{\partial}_x)$ is
the velocity operator along the $x$-direction.
From (\ref{eq:heisenberg}) the well known continuity equation $\partial_t \rho+\partial_x J=0$ is obtained.\\
In an analogous manner, the time evolution of the spin density
$S_\mu=\Psi^{\dag}\frac{\hbar}{2} \sigma_\mu\Psi$, $\mu=\{x,y,z\}$
($\sigma_\mu$ are the Pauli matrices) , can be derived:
\begin{equation}
\label{eq:no-continuity-spin}
\partial_t S_\mu=-\partial_xJ^s_\mu+\tau_\mu.
\end{equation}
Here $J^s_\mu=\Psi^{\dag}\frac{\hbar}{2} \sigma_\mu \hat{v}\Psi$
is the $\mu$-th component of the spin current density and
$\tau_\mu$ is the  the density of spin torque. When the
spin-dependent potential describes the Zeeman interaction between
the spin and a magnetic field along the direction $\hat{n}(x)$,
$V(x,\vec{\sigma})=\gamma(x) \hat{n}(x)\cdot\vec{\sigma}$, the
density of spin torque is explicitly given by
$\tau_\mu=\gamma(x)\Psi^{\dag}(\hat{n}(x)\times\vec{\sigma})_\mu\Psi=\frac{2\gamma(x)}{\hbar}(\hat{n}(x)\times
\vec{S})_\mu$. When $\gamma(x)=0$, the spin torque term vanishes
and a continuity equation is obeyed  by the spin currents,
$\partial_t \vec{S}+\partial_x\vec{J}^s=0$. In the case of a
system in which a magnetic central region $x\in [-a,a]$ is
connected to two external non-magnetic leads, under the stationary
 condition for the spin-current (i.e. $\partial_t \vec{S}=0$) the spin torque per unit
of area can be calculated by spatial integration of the equation
$\partial_xJ^s_\mu=\tau_\mu$ along the $x$-direction:
%\begin{equation}
%\label{eq:spin-current-gradient}
%\vec{J}^s(x\rightarrow +\infty)-\vec{J}^s(x\rightarrow -\infty)=\int^{+\infty}_{-\infty} dx \vec{\tau}.
%\end{equation}
\begin{equation}
\label{eq:spin-current-gradient}
\vec{J}^s_R-\vec{J}^s_L=\int^{+a}_{-a} dx \vec{\tau}= \vec{T},
\end{equation}
where $\vec{J}^s_{R/L}=\vec{J}^s(\pm\infty)$ is the value of the
current at infinity. The above result tells that the spin torque
per unit of area is due the difference of the spin currents on the
right and on the left of the magnetic scattering region. In the
following Section we analyze the spin torque transfer within a
scattering matrix approach.

\section{Density of spin current within the scattering approach}
\label{sec:scattering_formalism} In order to calculate the spin
torque through a magnetic region we employ the scattering matrix
formalism of Ref.[\onlinecite{buttiker92}] and consider a single
channel approximation to simplify the derivation.

The $S$-matrix connects the outgoing states to the incoming states
through the relation:
\begin{equation}
\label{eq:scattering_operator}
b^{\alpha}_{\sigma}=\sum_{\sigma'\beta}S^{\alpha
\beta}_{\sigma\sigma'}a^{\beta}_{\sigma'},
\end{equation}
where $b^{\alpha}_{\sigma},a^{\alpha}_{\sigma}$ are the scattering
operators for the outgoing and incoming states,  $\sigma=\pm$ is
the spin index, $\alpha=\{left, right\}$ represents the lead
index, while the $S$-matrix must be unitary, i.e. $S^{\dag}S=1$.
The quantum field representing the electron state in the lead
$\alpha$ can be expressed in terms of the scattering operators as
follows:
\begin{eqnarray}
&&\Psi_{\alpha}(x,t)=\sum_{\sigma}\int dE \rho_{\alpha}(E)
\exp\Bigl[-i\frac{E}{\hbar}t\Bigl]|\sigma \rangle\times\\\nonumber
&&[e^{i k x}a^{\alpha}_{\sigma}(E)+e^{-i k
x}b^{\alpha}_{\sigma}(E)],
\end{eqnarray}
where $\rho_{\alpha}(E)=[\sqrt{2\pi \hbar v_{\alpha}(E)}]^{-1}$ is
the density of states of lead $\alpha$, $v_{\alpha}(E)$ is the
velocity of the electrons with wave vector $k=\sqrt{2mE}/\hbar$
and the velocity is chosen to have positive orientation for the
incoming states. The spin-current density
$\vec{J}^s=\Psi^{\dag}\frac{\hbar}{2} \vec{\sigma} \hat{v}\Psi$
can be calculated with the approach of Ref.\cite{buttiker92} and
the result is:
\begin{eqnarray}
\label{eq:density-spin-curr-buttiker-like}
&& J^s_{\mu,\alpha}=\frac{1}{4\pi}\int dE dE' \exp\Bigl[i\frac{(E-E')}{\hbar}t\Bigl]\times\\\nonumber
&& [a^{\alpha \dag}(E)\sigma_{\mu}a^{\alpha}(E')-b^{\alpha \dag}(E)\sigma_{\mu}b^{\alpha}(E')],
\end{eqnarray}
where the following spinorial representation has been introduced:
\begin{equation}
\label{eq:spinorial}
a^{\alpha}=\Bigl(\begin{array}{c}
                   a_{+}^{\alpha} \\
                   a_{-}^{\alpha}
                 \end{array}\Bigl),\\
\end{equation}
while $a^{\alpha \dag}=(a_{+}^{\alpha \dag}, a_{-}^{\alpha
\dag})$. Using the relation $\langle a_{\sigma}^{\alpha
\dag}(E)a_{\sigma'}^{\beta}(E')\rangle=\delta_{\alpha
\beta}\delta_{\sigma\sigma'}f_{\beta}(E)$, $f_{\beta}(E)$ being
the Fermi function of the lead $\beta$ and
Eq.(\ref{eq:scattering_operator}), the $\mu$-th component of the
spin density current $\langle J^s_{\mu,\alpha} \rangle$ in the
lead $\alpha$ can be written as \cite{sharma-brouwer03-spin-curr}:
\begin{eqnarray}
\langle J^s_{\mu,\alpha} \rangle=-\frac{1}{4\pi}\sum_{\beta}\int dE Tr\{S^{\alpha \beta \dag}(E)\sigma_{\mu}S^{\alpha \beta}(E) \}f_{\beta}(E),
\end{eqnarray}
while using ( \ref{eq:spin-current-gradient}) the spin torque per
unit of area $T_\mu=\int dx \tau_\mu$ is:
\begin{eqnarray}
\label{eq:torque_smatrix}
T_\mu=\frac{1}{4\pi}\sum_{\alpha\beta=L,R}\int dE Tr\{S^{\alpha
\beta \dag}(E)\sigma_{\mu}S^{\alpha \beta}(E) \}f_{\beta}(E).
\end{eqnarray}
When a dc voltage bias $V$ is applied to the external leads,
changing their chemical potentials in  $\mu_{l,r}=E_F\pm eV/2$,
the variation of the spin-torque with respect to the external
perturbation $w=eV/2$ can be written as $\delta_w
T_\mu=\partial_{w}T_\mu \delta w$ in the linear response regime.
It is determined by the torkance $\partial_{w}T_\mu$. Using
Eq.(\ref{eq:torque_smatrix}) and taking the zero temperature
limit, we explicitly have\cite{carva09}:
\begin{eqnarray}
\label{eq:torkance-scattering}
\partial_{w}T_\mu=\frac{1}{2\pi}Tr\{ \sigma_\mu (S^{21}S^{21\dag}-S^{12}S^{12\dag})\}\Bigl
|_{E=E_F}.
\end{eqnarray}
The torkance along the direction of the unit vector
$\hat{n}=(n_x,n_y,n_z)$ is obtained as
$\hat{n}\cdot\partial_{w}\vec{T}$.
Eq.(\ref{eq:torkance-scattering}) describes the magnetic answer of
the system to an external dc voltage.
%RC
Let us note that compared to previous
works\cite{slonc_spintorque,waintal2000} our calculation of the
scattering matrix includes also off-diagonal spin-flip reflection
and transmission amplitudes.

\section{All electrical generation of spin torque by quantum pumping}
\label{sec:pumping}

\begin{figure}
\centering
\includegraphics[scale=0.4]{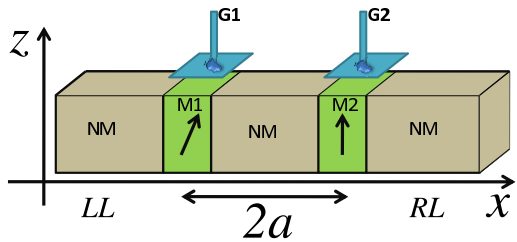}\\

\caption{Schematic representation of the NM/M/NM/M/NM system described in the main text.} \label{fig:fig1}
\end{figure}

In this section we discuss the generation of spin torque by means
of ac external gate voltages. In particular we generalize the
notion of quantum pumping of charges due to
Thouless\cite{thouless} to the spin torque. In a quantum pump a dc
particle current is generated by the ac adiabatic modulation of at
least two out-of-phase independent parameters of the system (e.g.
local magnetic fields or gate voltages)  in {\it absence of bias}.
In our calculation we will show that since pumping procedure in a
magnetic layered structure can generate spin currents other than
charge currents, a spin-torque is generated by the gradient of
spin current (see
Eqs.(\ref{eq:no-continuity-spin})-(\ref{eq:spin-current-gradient})).
In particular, we focus on the system of Fig.(\ref{fig:fig1}) in
which a microstructure made of a central non-magnetic-region (NM)
is connected to two external non-magnetic leads (LL and RL)
through magnetic layers (M1 and M2) (whose width is taken less
than the De Broglie wavelength). Applying the idea of pumping, we
modulate in time the barriers heights at the interface between the
M-NM regions by the top gates G1 and G2. In the presence of this
ac modulation the scattering matrix $S$ depends explicitly on time
and the relation between incoming and outgoing states is:
\begin{equation}
b^{\alpha}(t)=\sum_{\beta}\int dt'S^{\alpha
\beta}(t,t')a^{\beta}(t'),
\end{equation}
where the spin-indices are absorbed in the spinorial notation (see Eq.(\ref{eq:spinorial})). When the gates
are varied adiabatically in time, an instantaneous approximation
can be made, i.e. $S^{\alpha
\beta}(t,t')=\delta(t-t')S^{\alpha\beta}(t)$, and thus
$b^{\alpha}(t)=\sum_{\beta}S^{\alpha \beta}(t)a^{\beta}(t)$. In
particular, since the time dependence of the scattering matrix
$S(t)$ is induced by two external parameters of the form
\begin{eqnarray}
&&X_1(t)=X_1^0+X_1^{\omega}\sin(\omega t)\\\nonumber
&&X_2(t)=X_2^0+X_2^{\omega}\sin(\omega t+\varphi),
\end{eqnarray}
then $S(t)=S(X_1(t),X_2(t))$. In particular, when the amplitude of
the ac parameters is small, i.e. $X_{1,2}^{\omega}\ll
X_{1,2}^{0}$, the scattering matrix can be expanded as follows:
\begin{equation}
\label{eq:exp_S} S^{\alpha \beta}(t)\simeq S_0^{\alpha
\beta}+\sum_{\eta= \pm1}s_{\eta}^{\alpha\beta}e^{i\eta \omega t},
\end{equation}
where $\omega=2\pi\nu$ is the angular frequency of the adiabatic
modulation and the matrices $s_{\eta}^{\alpha\beta}$ are given by
\begin{equation}
\label{eq:small_s}
s_{\eta}=-\frac{i\eta}{2}\Bigl[X_1^{\omega}(\partial_{X_1}S)_0+X_2^{\omega}e^{i\eta\varphi}(\partial_{X_2}S)_0\Bigl].
\end{equation}
The Fourier transform of (\ref{eq:exp_S}) is then:
\begin{equation}
\label{eq:ft_exp_S} S^{\alpha \beta}(E)= 2\pi[ S_0^{\alpha
\beta}\delta(E)+\sum_{\eta=
\pm1}s_{\eta}^{\alpha\beta}\delta(E+\eta \omega)].
\end{equation}
Consequently, the relation between the outgoing and incoming
states in the Fourier space takes the following form:
\begin{equation}
\label{eq:b(E)}
b^{\alpha}(E)=\sum_{\beta}\Bigl[S_0^{\alpha \beta}a^{\beta}(E)+\sum_{\eta}s^{\alpha \beta}_{\eta}a^{\beta}(E+\eta \omega)\Bigl].
\end{equation}
Using Eq.(\ref{eq:b(E)}) in
(\ref{eq:density-spin-curr-buttiker-like}) the $\mu$-component of
the spin torque per unit of area is given by:
\begin{equation}
T_\mu=\frac{1}{4\pi}\sum_{\eta\alpha\beta}\int dE
Tr\{s^{\alpha\beta \dag}_{\eta}\sigma_{\mu}s^{\alpha\beta
}_{\eta}\}f_{\beta}(E+\eta\omega).
\end{equation}
 Since no bias is present between the leads $f_{\beta}(E)=f(E)$,
and in the zero temperature limit, the $\mu$-th component of
spin-torque per unit area to leading order in the adiabatic frequency $\omega$ is:
\begin{equation}
T_\mu=-\frac{\hbar\omega}{4\pi}\sum_{\alpha\beta\eta}\eta Tr\{\sigma_{\mu}s^{\alpha\beta}_{\eta}s^{\alpha\beta \dag}_{\eta}\},
\end{equation}
(all terms independent from the external perturbation have been
dropped). Using (\ref{eq:small_s}) this can be rewritten in terms
of the parametric derivatives of the scattering matrix as :
\begin{eqnarray}
\label{eq:torque_pump} T_{\mu}&=&\frac{\hbar\omega
X_1^{\omega}X_2^{\omega}\sin(\varphi)}{8\pi}\sum_{\alpha\beta}Tr\{A_{\mu}^{\alpha\beta}+A_{\mu}^{\alpha\beta\dag}\},
\end{eqnarray}
where we introduced the quantity
$A_{\mu}^{\alpha\beta}=i(\partial_{X_2}S^{\alpha\beta\dag})_0\sigma_{\mu}(\partial_{X_1}S^{\alpha\beta})_0$.
The torque generated in the direction of the unit vector $\hat{n}$
is given by $T_{||}=\hat{n}\cdot\vec{T}$.
Eq.(\ref{eq:torque_pump}) represents the torque pumped in a
magnetic system and in principle could be very different from the
one induced by a dc voltage. In the following we perform the
explicit calculation for the system of Fig. 1.

\section{Spin torque in M|NM|M systems}
\label{sec:calculation} Here we analyze the spin torque generated
in the system of Fig. 1 in the following two situations: (i) in
the presence of a dc voltage applied to the external leads; (ii)
in the absence of bias and using a quantum pumping procedure.
%===================================================================fig1
%=======================================================================
The central NM region is connected to two external leads via the
thin magnetic layers M1 and M2 and without loss of generality we
choose the magnetization of M2 parallel to the $z$-axis, while the
magnetization of the region M1 has arbitrary direction. The
distance between the two magnetic barriers is taken $2a$ and their
height is controlled by the the top gates G1 and G2. A minimal
model for the system above is given by the following Hamiltonian:
\begin{equation}
H=-\frac{\hbar^2}{2m}\partial^2_x+V_m(x,\vec{\sigma})+U(x),
\end{equation}
where $U(x)=g_1\delta(x+a)+g_2\delta(x-a)$ is the barriers
potential at the interface between M-NM region, while
$V_m(x,\vec{\sigma})=\gamma_l\delta(x+a)\hat{n}\cdot\vec{\sigma}+\gamma_r\delta(x-a)\sigma_z$
represents the magnetic interaction between the conduction
electrons spin and the magnetic moment of the layers M1 and M2.
The scattering problem for the  Hamiltonian above can be easily
solved by imposing the appropriate boundary conditions on the
wavefunctions and their derivatives at the interfaces (see, for
instance, Ref.[\onlinecite{chen04}]) and from the knowledge of the
scattering matrix the spin torque $T_{\mu}$ can be calculated
using the equations of the previous sections.
%RC: mettere l'espressione di t ed r se possibile
%RC
The following consideration is in order here: The presence of two
magnetic layers has the effect of allowing for multiple scattering
of the electrons between them, which gives rise to an explicit
asymmetry of the spin current and produces a finite spin-torque
effect.

 In the following we employ  the adimensional quantities: $z_1=\frac{2m
g1}{\hbar^2k_F}$, $z_2=\frac{2m g2}{\hbar^2k_F}$,
$\Gamma_R=\frac{2m \gamma_r}{\hbar^2k_F}$, $\Gamma_L=\frac{2m
\gamma_l}{\hbar^2k_F}$, where $k_F$ is the Fermi wavelength, $m=\alpha m_e$
is the effective electron mass, while $\alpha$ represents the ratio between the effective and the bare electron mass $m_e$. The distances are made
adimensional by multiplying by $k_F$, i.e. $a\rightarrow k_F a$.

\subsection{Spin torque by means of dc voltage bias}
\begin{figure}
\centering
\includegraphics[scale=0.8]{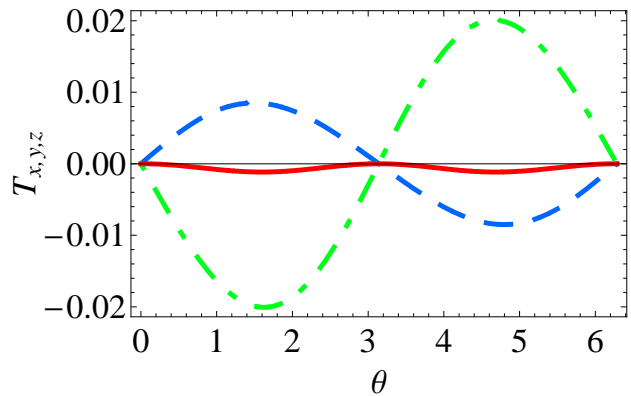}\\

\caption{Torque $T_x$ (dashed line), $T_y$ (dashed-dotted line),
$T_z$ (full line) in unit of $eV/2$ computed as a function of the
angle $\theta$ between M1 and M2. The parameters have been fixed
as follows: $a=2$, $z_1=0.5$, $z_2=1$, $\Gamma_L=0.1$,
$\Gamma_R=1$. When the magnetization directions of the layers M1
and M2 are parallel or anti-parallel the spin torque vanishes.}
\label{fig:fig2}
\end{figure}

Here we present the results of the spin torque per unit of area generated by a small dc voltage bias $V$ applied to the external leads.
 In the following we use polar coordinates for the magnetic moment $\vec{M_1}=M_1 \hat{n}$,
 $\hat{n}=(\sin(\theta)\cos(\phi),\sin(\theta)\sin(\phi),\cos(\theta))$,
 and assume that the momenta $\vec{M_1}$ and $\vec{M_2}$ lie in the plane $x-z$. In this case we have $\hat{n}=(\sin(\theta),0,\cos(\theta))$, where the $\theta$ represents the angle between $\vec{M_1}$ and $\vec{M_2}$.\\
In Fig.(\ref{fig:fig2}) the three components of the torque (in
unit of $eV/2$) are shown as a function of angle $\theta$ between
the magnetic momenta of the regions M1 and M2 and by fixing the
remaining parameters as follows: $a=2$, $z_1=0.5$, $z_2=1$,
$\Gamma_L=0.1$, $\Gamma_R=1$.

 Similarly to other
works\cite{waintal2000}, we find that all the torque components
vanish when M1 and M2 are parallel or anti-parallel. Furthermore,
the $z$-component $T_z$ of the torque is very small compared to
$T_x$ and $T_y$, while the component perpendicular to the plane of
the magnetic moment $T_y$ is the strongest. This component of the
torque follows a $\sin \theta$ behavior for the specific values of
magnetic interactions considered in the figure ($\Gamma_R \gg
\Gamma_L$). In fact when the non-spin-polarized electrons are
incident on the magnetic layer M2, spin-filtering removes the
component of the spin angular momentum perpendicular to the layer
moment from the current acting as a polarizer along the $z$
direction. Thus the polarized electrons scattered off M2 and
incident on M1 generate an effective torque on the layer momentum
M1 proportional to the $\sin \theta$.
%RC: mettere il perche'
%RC
The negative sign of the spin torque component $T_y$ is due to the
fact that electrons with spins parallel to the moment of the
magnetic region M2 have a larger transmission probability compared
to those antiparallel.

%===================================================================fig2
\begin{figure}
\centering
\includegraphics[scale=0.8]{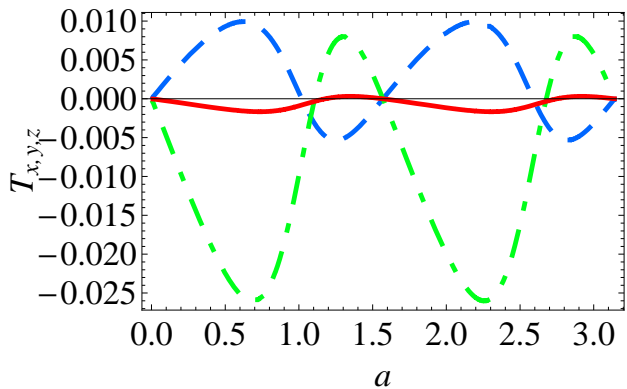}\\

\caption{Torque $T_x$ (dashed line), $T_y$ (dashed-dotted line), $T_z$ (full line) in unit of $eV/2$ computed as a function of the semi-distance $a$ between the gates G1 and G2. The parameters have been fixed as follows: $\theta=1.5$, $z_1=0.5$, $z_2=1$, $\Gamma_L=0.1$, $\Gamma_R=1$. Notice that the torque $T_y$, i.e. the torque in the direction perpendicular to the plane $x-z$ presents large oscillations compared with the in-plane components.} \label{fig:fig3}
\end{figure}

%=======================================================================
In Fig.(\ref{fig:fig3}) the components of the torque are shown as
a function of the semi-distance $a$ between the gates G1 and G2
and by fixing the remaining parameters as follows: $\theta=1.5$,
$z_1=0.5$, $z_2=1$, $\Gamma_L=0.1$, $\Gamma_R=1$. For a mesoscopic
system characterized by a De Broglie wavelength
$\lambda\approx20-30$nm the adimensional distance $a=1$ would
correspond to $3.18-4.78$nm. As shown in the figure, the torque
presents a characteristic oscillatory behavior with the
semi-length of the central region $a$.
%RC
These oscillations can be regarded as a quantum-size effect. They
reflect the perfect ballistic regime of electron transport through
the spin valve. The physical mechanism behind the oscillations is
the interference effect of the electrons propagating across the
non-magnetic-magnetic interface from the right lead to the left
lead and electrons propagating backwards.  The particular value of
the oscillation follows from the values of the spin-dependent
Fermi wavevector.

%===================================================================fig3
\begin{figure}
\centering
\includegraphics[scale=0.8]{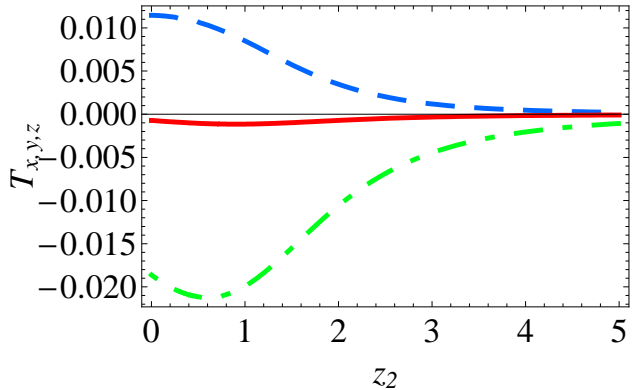}\\

\caption{Torque $T_x$ (dashed line), $T_y$ (dashed-dotted line),
$T_z$ (full line) in unit of $eV/2$ computed as a function of the
non-dimensional barrier strength $z_2$ controlled by G2. The
parameters have been fixed as follows: $\theta=1.5$, $z_1=0.5$,
$a=2$, $\Gamma_L=0.1$, $\Gamma_R=1$.
%Notice that the all the
%torque component induced by the gradient of the spin currents
%vanishes when the transmission probability of the electron becomes
%strongly suppressed by the strength of the barrier.
}
\label{fig:fig4}
\end{figure}

\begin{figure}
\centering
\includegraphics[scale=0.8]{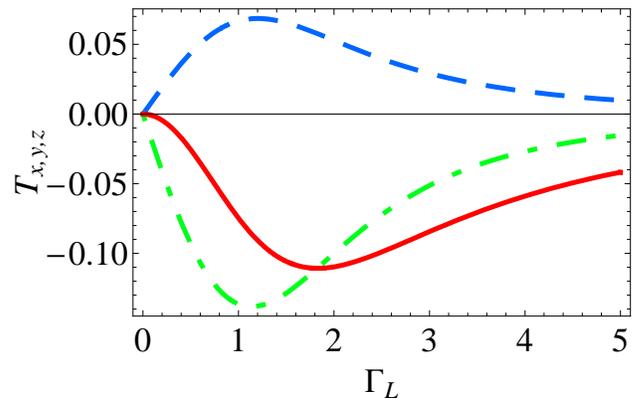}\\

\caption{Torque $T_x$ (dashed line), $T_y$ (dashed-dotted line),
$T_z$ (full line) in unit of $eV/2$ computed as a function of
$\Gamma_L$ controlled by the magnetic momentum in the region M1.
The remaining parameters have been fixed as follows: $\theta=1.5$,
$z_1=0.5$, $z_2=0.8$, $a=2$, $\Gamma_R=1$.
%Notice that the torque
%components present maximum values for different strength of the
%magnetic momentum of the region M1.
} \label{fig:fig5}
\end{figure}

%=======================================================================
In Fig.(\ref{fig:fig4}) we plot $T_{x,y,z}$ as a function of
the strength $z_2$ of the right barrier (controllable via the gate
G2) by fixing the other parameters as:
 $\theta=1.5$, $z_1=0.5$, $a=2$, $\Gamma_L=0.1$, $\Gamma_R=1$.
The torque of course depends crucially on the transparency of the
barrier and thus the components $T_{x,y}$ decreases with
increasing barrier height becoming comparable to $T_z$.  An
interesting feature appears by increasing the Zeeman interaction
$\Gamma_L$ as shown in Fig.(\ref{fig:fig5}): for values of
$\Gamma_L$ above $0.2$ the $z$-component of the torque $T_z$
starts to assume relevant values. In fact the magnetic interaction
becomes more effective in aligning the electron spins along the
magnetic moment of the layer M1 thus reducing the contribution to
the gradient of spin current.
% by setting the remaining parameters
% as: $\theta=1.5$, $z_1=0.5$, $a=2$, $\rho=0.1$, $\Delta=1$.
%===================================================================fig4
%=======================================================================
%===================================================================fig5
%=======================================================================

\subsection{Spin torque generation by means of adiabatic quantum pumping}
An alternative way to generate a spin torque is the adiabatic
quantum pumping technique. As explained above we keep the two
external leads at the same chemical potential and in order to
generate a spin current we modulate out-of-phase in time the two
barrier heights $z_1$ and $z_2$ by the gate voltages G1 and G2:
\begin{eqnarray}
&&z_1(t)=z_1^0+z^{\omega}_1\sin(\omega t)\\\nonumber
&&z_2(t)=z_2^0+z^{\omega}_2\sin(\omega t+\varphi).
\end{eqnarray}
Thus by using (\ref{eq:torque_pump}) the $\mu-th$ component of the
torque in the adiabatic regime is given by:
\begin{eqnarray}
T_{\mu}&=&\frac{\hbar\omega
z_1^{\omega}z_2^{\omega}\sin(\varphi)}{8\pi}\sum_{\alpha\beta}Tr\{A_{\mu}^{\alpha\beta}+A_{\mu}^{\alpha\beta\dag}\}.
\end{eqnarray}
In the weak pumping regime (i.e. $z_i^0 \gg z_i^{\omega}$)
considered here, $T_\mu$ presents a $\sin(\varphi)$
behavior with respect to the pumping phase $\varphi$ and thus in the following analysis we set $\varphi=\pi/2$
and measure the spin torque in unit of $\hbar\nu/2$.\\
Concerning the frequencies of the pump, we can safely consider as
adiabatic the frequencies for which $\hbar \omega$ is smaller than
the first gap in the electron energy spectrum such that the system
lies in its ground state\cite{thouless}. For simplicity, by
considering the energy spectrum of a particle confined in a box,
we easily find that the threshold frequency is $\tilde{\nu}\sim
\frac{h}{2m^\star L^2}$ , where $L$ is the length of the system,
while $m^\star=\alpha m_e$ is the effective mass of the electrons.
Thus for a system with $L=1.5\mu$m and effective mass ratio
$\alpha=0.023$, $\tilde{\nu}\approx 43.4$ GHz which is consistent
with the values of the frequencies of quantum pumping in quantum
dots. For a smaller system characterized by $L=0.5\mu$m and
$\alpha=0.023$ we obtain $\tilde{\nu}\approx 396.5$ GHz which is a
more convenient value to achieve in experiments.

\begin{figure}
\centering
\includegraphics[scale=0.8]{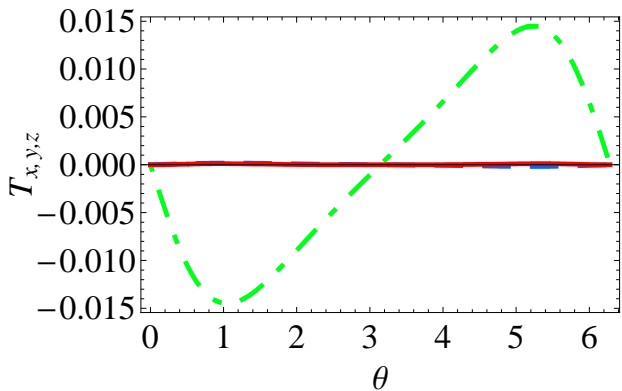}\\

\caption{Torque $T_x$ (dashed line), $T_y$ (dashed-dotted line),
$T_z$ (full line) in unit of $\hbar\nu/2$ computed as a function
of $\theta$ between M1 and M2. The remaining parameters have been
fixed as follows: $z_1^{\omega}=z_2^{\omega}=0.2$, $\Gamma_L=0.9$,
$z_1^0=0.5$, $z_2^0=0.3$, $a=1$, $\Gamma_R=1$.
%Notice that in the
%present configuration the only relevant component of the spin
%torque is $T_y$. When the magnetization directions of the region
%M1 and M2 are parallel or anti-parallel $T_y$ vanishes.
}
\label{fig:fig6}
\end{figure}

\begin{figure}
\centering
\includegraphics[scale=0.8]{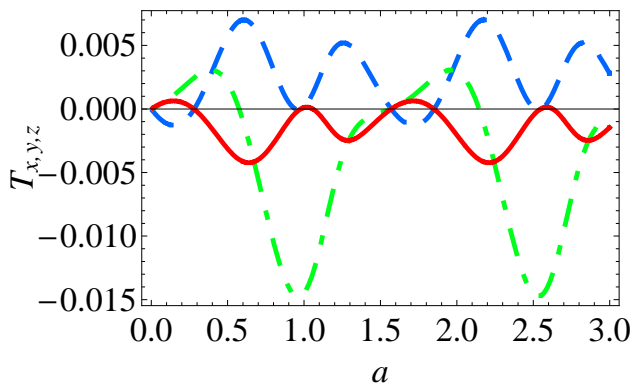}\\

\caption{Torque $T_x$ (dashed line), $T_y$ (dashed-dotted line), $T_z$ (full line) in unit of $\hbar\nu/2$ computed as a function of the semi-distance $a$ between the gates G1 and G2. The remaining parameters have been fixed as follows: $z_1^{\omega}=z_2^{\omega}=0.2$, $\Gamma_L=0.9$, $z_1^0=0.5$, $z_2^0=0.3$, $\Gamma_R=1$, $\theta=1$.} \label{fig:fig7}
\end{figure}

\begin{figure}
\centering
\includegraphics[scale=0.8]{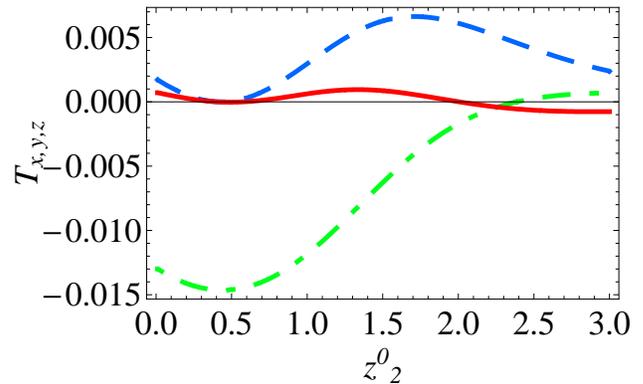}\\

\caption{Torque $T_x$ (dashed line), $T_y$ (dashed-dotted line), $T_z$ (full line) in unit of $\hbar\nu/2$ computed as a function of the static part of the non-dimensional barrier strength $z_2^0$ controlled by G2. The remaining parameters have been fixed as follows: $z_1^{\omega}=z_2^{\omega}=0.2$, $\Gamma_L=0.9$, $z_1^0=0.5$, $a=1$, $\Gamma_R=1$, $\theta=1$.} \label{fig:fig8}
\end{figure}

\begin{figure}
\vspace{1.0cm}
\centering
\includegraphics[scale=0.8]{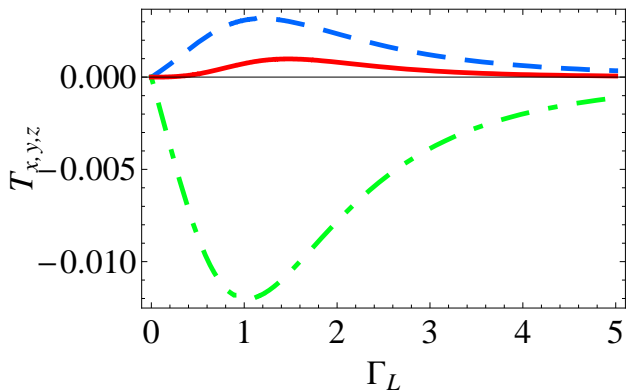}\\

\caption{Torque $T_x$ (dashed line), $T_y$ (dashed-dotted line), $T_z$ (full line) in unit of $\hbar\nu/2$ computed as a function of  $\Gamma_L$ controlled by the magnetic momentum in the region M1. The remaining parameters have been fixed as follows: $z_1^{\omega}=z_2^{\omega}=0.2$, $z_2^0=1$, $z_1^0=0.5$, $a=1$, $\Gamma_R=1$, $\theta=1$.} \label{fig:fig9}
\end{figure}

In Fig.(\ref{fig:fig6}) we plot the components $T_\mu$ as a
function of the angle $\theta$ between the magnetization M1 and M2
and by fixing the other parameters as follows:
$z_1^{\omega}=z_2^{\omega}=0.2$, $\Gamma_L=0.9$, $z_1^0=0.5$,
$z_2^0=0.3$, $a=1$, $\Gamma_R=1$.
%RC
Compared to the case of dc bias, the $\theta$ dependence of the
torque is not a simple $\sin \theta$ form because the strength of
the magnetic interactions $\Gamma_L$ and $\Gamma_R$ is comparable
and thus both $\cos \theta$ and $\sin \theta$ terms contribute.
 Moreover as in the previous analysis, while the in-plane component
of the spin torque $T_x$ and $T_z$ assume very small values in the
parameter region considered, the $T_y$ component presents relevant
values. Indeed, by setting $\theta \approx 1$ and for pumping
frequencies $\nu$ of the order of 10-30 GHz we have a torque
component $T_y \approx 0.05-0.15$$\mu$eV per unit of area, which
is of the same order of magnitude obtained in a similar system
under dc bias (see Ref.[\onlinecite{kalitsov09}]).
%===================================================================fig6
%=======================================================================
In Fig.(\ref{fig:fig7}) we show the spin torque components
$T_{x,y,z}$ as a function of the semi-distance $a$ between the
gates G1 and G2, for the remaining parameters:
$z_1^{\omega}=z_2^{\omega}=0.2$, $\Gamma_L=0.9$, $z_1^0=0.5$,
$z_2^0=0.3$, $\Gamma_R=1$, $\theta=1$. Compared to the case in
which the spin-torque is generated by an external dc bias, more
harmonics are present in the torque dependence on $a$. This
behavior can be qualitatively explained by the fact that the
torque generated by the pumping procedure is related to the
parametric derivatives of the scattering matrix.
%===================================================================fig7
%=======================================================================
In Fig.(\ref{fig:fig8}) we show the spin torque components
$T_{x,y,z}$ as a function of the static part of the adimensional
barrier strength $z_2^0$ controlled by G2, while the remaining
parameters have been fixed as follows:
$z_1^{\omega}=z_2^{\omega}=0.2$, $\Gamma_L=0.9$, $z_1^0=0.5$,
$a=1$, $\Gamma_R=1$, $\theta=1$. As shown, for increasing values
of the static barrier strength $z_2^0$ the torque pumped in the
system becomes vanishes. Indeed, the reduction of the spin current
from the external leads reduces the gradient of the current and
thus the spin torque is strongly suppressed. In the case of
pumping, the suppression of the torque by increasing the barriers
height may be in general stronger compared to the dc case.
%===================================================================fig8
%=======================================================================
In Fig.(\ref{fig:fig9}) we show the spin torque components
$T_{x,y,z}$ as a function of the Zeeman interaction $\Gamma_L$ for
the other parameters: $z_1^{\omega}=z_2^{\omega}=0.2$, $z_2^0=1$,
$z_1^0=0.5$, $a=1$, $\Gamma_R=1$, $\theta=1$. A behavior of the
torque similar to the one presented in Fig.(\ref{fig:fig5}) is
found.
%RC
The magnetic interaction of the electron spins with the magnetic
moment of M1 causes a suppression of the torque above
$\Gamma_L\approx 1.7$ thus acting like an energy barrier for the
electrons.
%This statement can be easily demonstrated exploiting a simplified
%model of domain wall by performing a position-dependent unitary
%transformation\cite{dugaev07}. In such a case the local energy
%(due to the change in the quantization axis of the electron spin)
%is a function of the inverse of the length of the magnetic region
%and thus produces a barrier effect as stronger as the magnetic
%region is small compared to the De Broglie wavelength.
%===================================================================fig9
%======================================================================
\section{Conclusions}
Within the scattering matrix approach we studied the spin torque
generated by dc or ac external perturbations acting on a
multilayered system consisting of a sequence of
magnetic/nonmagnetic regions. In particular, we studied the
spin-torque of a magnetic-nonmagnetic-magnetic trylayer connected
to metallic nonmagnetic leads.
%RC
We have focused on the effect of spin-filtering as the mechanism
for current induced torque, i.e. the difference in the
transmission and reflection probabilities for electrons with spins
parallel and antiparallel to the moments of the magnetic layers.
As a source of spin-dependent scattering we included also the
spin-flip mechanism (i.e. the off diagonal elements of the
scattering matrix).
 The spin torque
generated by the application of a dc bias to the external leads
was analyzed as a function of the relative orientation $\theta$
between the magnetic moments of the magnetic regions, the length
$2a$ of the nonmagnetic spacer, the barriers transparencies and
the Zeeman couplings. From the analysis we observed that the
spin-torque vanishes when the magnetic moments of M1 and M2 are
parallel or antiparallel, while an oscillating behavior of the
spin-torque was observed as a function of $a$, indicating a
quantum size effect. When the transparencies of the tunnel
barriers at the interfaces between the magnetic/nonmagnetic layers
are lowered, a strong suppression of the spin torque is observed
due to the reduction of the spin fluxes through the barriers.
Similar effects were found as a function of the strength of the
magnetic interaction $\Gamma_L$.

%RC
As an alternative to considering external dc bias, we proposed a
current induced spin torque based on quantum pumping, generalizing
the original idea of Thouless for the charges \cite{thouless}. We
formulated a scattering matrix formalism of the spin torque
pumping. By modulating in time the strength of the two
out-of-phase tunnel barriers at the interface between the
magnetic/nonmagnetic layers, spin polarized currents were produced
in the leads and a gradient of spin current responsible for the
spin torque was generated. By studying the spin torque components
in this case we observed different signatures of the magnetic
reversal mechanism due to the pumping procedure. For instance,
within the weak pumping limit additional harmonics were observed
in the oscillations of $T_{\mu}$ \textit{vs} the length of the
nonmagnetic region $2a$ and a stronger suppression of the spin
torque at increasing the barrier heights.
%Moreover the behavior of $T_{\mu}$ as a
%function of the relative orientation of the magnetic moments was
%found to deviate from a simple $\sin \theta$ form.
Thus the pumping mechanism could offer an additional tool to study
experimentally the magnetic response of metallic heterostructures.

Experimentally, current induced spin torque has been realized in
Co/Cu/Co sandwich structures\cite{cocuco1,cocuco2} and in Py/Cu/Py
heterostructures\cite{pycupy}. Our proposal can also be realized
by using two ferromagnetic EuS barriers coupled to Al leads and
separated by an $Al_2O_3$ spacer of few tens of nanometers by
modifying, for instance, the experimental setup studied in
Ref.[\onlinecite{guo-xing-miao09prl}].
%on the basis of the theory
%presented in Ref.[\onlinecite{saffarzadeh03}], the heterostructure
%being  of the form (10 nm) Al/(1.5 nm) EuS/(2.25 nm) $Al_2O_3$/(3
%nm) EuS/(10 nm) Al.

%RC: la borsa postdoc non va menzionata negli acknowledgment, basta mettere l'affiliazione.
%\section*{Acknowledgements}
%One of the authors F.R. gratefully acknowledges financial support from C.N.I.S.M. through a post-doctoral fellowship.
% ----------------------------------------------------------------
\bibliographystyle{prsty}

%====================================================================
\end{document}